\documentclass[12pt]{iopart}
\usepackage{graphicx}
\usepackage{lipsum} 

\begin{document}
\title[The reduction of entropy uncertainty for qutrit system under non-Markov noisy environment]{The reduction of entropy uncertainty for qutrit system under non-Markov noisy environment}
\author{XiongXu$^1$, MaofaFang$^1$\footnote{Synergetic Innovation Center for Quantum Effects and Applications, Key Laboratory of Low-dimensional Quantum Structures and Quantum Control of Ministry of Education, School of Physics and Electronics, Hunan Normal University, Changsha, 410081, P.R. China. } }
\address{$^1$ Synergetic Innovation Center for Quantum Effects and Applications, Key Laboratory of Low-dimensional Quantum Structures and Quantum Control of Ministry of Education, School of Physics and Electronics, Hunan Normal University, Changsha, 410081, P.R. China.}
\ead{mffang@hunnu.cn}
\begin{abstract}
In this paper, we explore the entropy uncertainty for qutrit system under non-Markov noisy environment and discuss the effects of the quantum memory system and the spontaneously generated interference (SGI) on the entropy uncertainty in detail. The results show that, the entropy uncertainty can be reduced by using the methods of quantum memory system and adjusting of SGI. Particularly, the entropy uncertainty can be decreased obviously when both the quantum memory system and the SGI are simultaneously applied.
\end{abstract}
\maketitle
\ioptwocol
\section{Introduction}
\label{intro}
The uncertainty principle, originally proposed by Heisenberg, clearly illustrates the difference between classical and quantum mechanics\cite{Heisenberg.1927}. Initially, the expression for the Heisenberg uncertainty principle with respect to the position $x$ and momentum $p_{x}$ is formulated as $\Delta x \cdot \Delta p_{x}\geq \hbar/2$, and later was generalized by Kennard and Robertson\cite{Robertson.1929} to two arbitrary observables R and S with a standard deviation
\begin{eqnarray}
   \Delta R \cdot \Delta S \geq \frac{|\langle[R,S]\rangle|}{2},
\end{eqnarray}
where the variance $\Delta Q = \sqrt{\langle Q^2\rangle - {\langle Q \rangle}^2}$(Q is an arbitrary observable).
$\langle \cdot \rangle$ is the expectation of the observable in a quantum system $\rho$, and $[R,S]$ is the commutator. However, because the lower bound of the above uncertainty relation is state dependent, the inequality will be trivial if the expectation value of the commutator $[R,S]$ is equal to zero and the variances only contain the second-order statistical moments of the quantum fluctuations of measurements. To remove this pitfall and to obtain a more general form which is independent of the quantum state, Deutsch recast that the uncertainty can be quantified by the Shannon entropy instead of the standard deviation. After a further improvement and proof, the classical entropy uncertainty relation is defined by\cite{MaassenH.UffinkJ.B.M.:.1988}
\begin{eqnarray}
   H(R)+H(S)\geq \log_{2}\frac{1}{c},
\end{eqnarray}
where $H(R)(H(S))$ is the Shannon entropy of
 the observable R(S), and $c=max_{i,j}{|\langle r_{i}|s_{j}\rangle |}^2$ is  the maximum overlap of eigenvectors ${|r_{i}\rangle}$ and ${|s_{j}\rangle}$ corresponding to a pair of incompatible observables R and S, respectively. More recently, a new entropy uncertainty relation has been derived by Renes and Boileau\cite{RenesJ.M.BoileauJ.C..2009,RenesJ.M.BoileauJ.C..2008} and later proved by Berta et.al.\cite{Berta.2010,LiC.F.XuJ.S.XuX.Y.LiK.GuoG.C.:.2011}, which is named quantum memory assisted the entropy uncertainty relation(QMA-EUR)
\begin{eqnarray}
   S(R|B)+S(Q|B)\geq \log_{2}\frac{1}{c}+S(A|B),
\end{eqnarray}
where $S(A|B)=S(\rho_{AB})-S(\rho_{B})$ is the conditional entropy and $S(\rho)=-Tr(\rho\log_2\rho)$ is the von Neumann entropy. $S(X|B)$ with $X\in{\{R,Q\}}$ is the conditional entropy of the post-measurement state $\rho_{XB}=\sum_i(|\psi_i\rangle\langle\psi_i|\otimes I)\rho_{AB}(|\psi_i\rangle\langle\psi_i|\otimes I)$ after particle A is measured by X, where ${|\psi_i\rangle}$ is the eigenstate of the observable $X$ and $I$ is an identity operator in the Hilbert space of particle B.
Generally, one can see the new uncertainty relation in term of the uncertainty game between two players, Alice and Bob.
Firstly Bob sends a particle A to Alice, which may entangle with his quantum memory B in general.
Secondly, Alice measures either R or S and notes her outcome.
Finally, Alice announces her measurement choice to Bob and Bob's task is to minimize his uncertainty about Alice's measurement outcome.

Recently, the improvement of this new entropic uncertainty principle as well as its dynamics have attracted increasing attention and there are potential applications such as for witnessing entanglement\cite{Berta.2010,LiC.F.XuJ.S.XuX.Y.LiK.GuoG.C.:.2011,Prevedel.2011,Hall.2012,Coles.2014,Zou.2014}, teleportation\cite{Hu.2012}
quantum phase transitions\cite{Nataf.2012} and cryptography\cite{Tomamichel.2012,Konig.2012,Dupuis.2015}.
In this paper, we explore the behavior of the entropic uncertainty for a qutrit system under non-Markov noisy environment.
In fact, any systems are essentially open and unavoidably interact with their surrounding environment, which induces decoherence or dissipation phenomena\cite{ZouH.M.FangM.F..2013,Xu.2013}.
People usually tend to pay attention to the entropy uncertainty of qubit system.
Feng et al explored the quantum-memory-assisted entropic uncertainty relation under noises and found that the unital noises only increase the entropic uncertainty, whereas the amplitude-damping non-unital noises may reduce the entropic uncertainty in the long-time limit\cite{Xu.2012}.
Fan et al studied the relations between the quantum-memory-assisted entropic uncertainty principle, quantum teleportation and entanglement witness\cite{Hu.2012b}.
Zhang et al studied the quantum-memory-assisted entropic uncertainty principle for the qubit system in non-Markovian environments\cite{Zhang.2018}. However, there are few researches on the entropy uncertainty of high-dimensional quantum system.
Quantum systems with a higher dimension may have advantages over ones with lower dimensional  during quantum information processing \cite{Walborn.2006,Bourennane.2001} since they provide higher channel capacities, more secure cryptography as well as superior quantum gates \cite{Cerf.2002,Durt.2003}.
Guo et al studied the entropic uncertainty relation for a qutrit system under decoherence\cite{Guo.2018}.
Zhang et al studied entropic uncertainty relation of a two-qutrit Heisenberg spin model in nonuniform magnetic fields and its dynamics under intrinsic decoherence\cite{Zhang.2018b}.
The entropy uncertainty of qutirt in non-Markovian environments has not been reported, so our work further enriches the study of the entropy uncertainty.

The paper is organized as follows. In Sec. 2, we present our physical model which consider a V-type three-level atom coupling to a reservoir of electromagnetic radiation modes. In Sec. 3, the entropy uncertainty with a memory system is investigated in  non-Markovian regimes.
In Sec. 4, we briefly discuss entropy uncertainty with SGI under non-Markovian environment.
Finally, we summarize our work in Sec. 5.
\section{Physical Model}
\label{sec:1}
We will consider a three-level atom that is in V-type atomic configuration, with states denoted $|0\rangle,\ |1\rangle$ and $|2\rangle$, coupling to a reservoir of electromagnetic radiation modes which is to be at zero temperature.
\begin{figure}[htpb]
\begin{center}
  \includegraphics[width=0.33\textwidth]{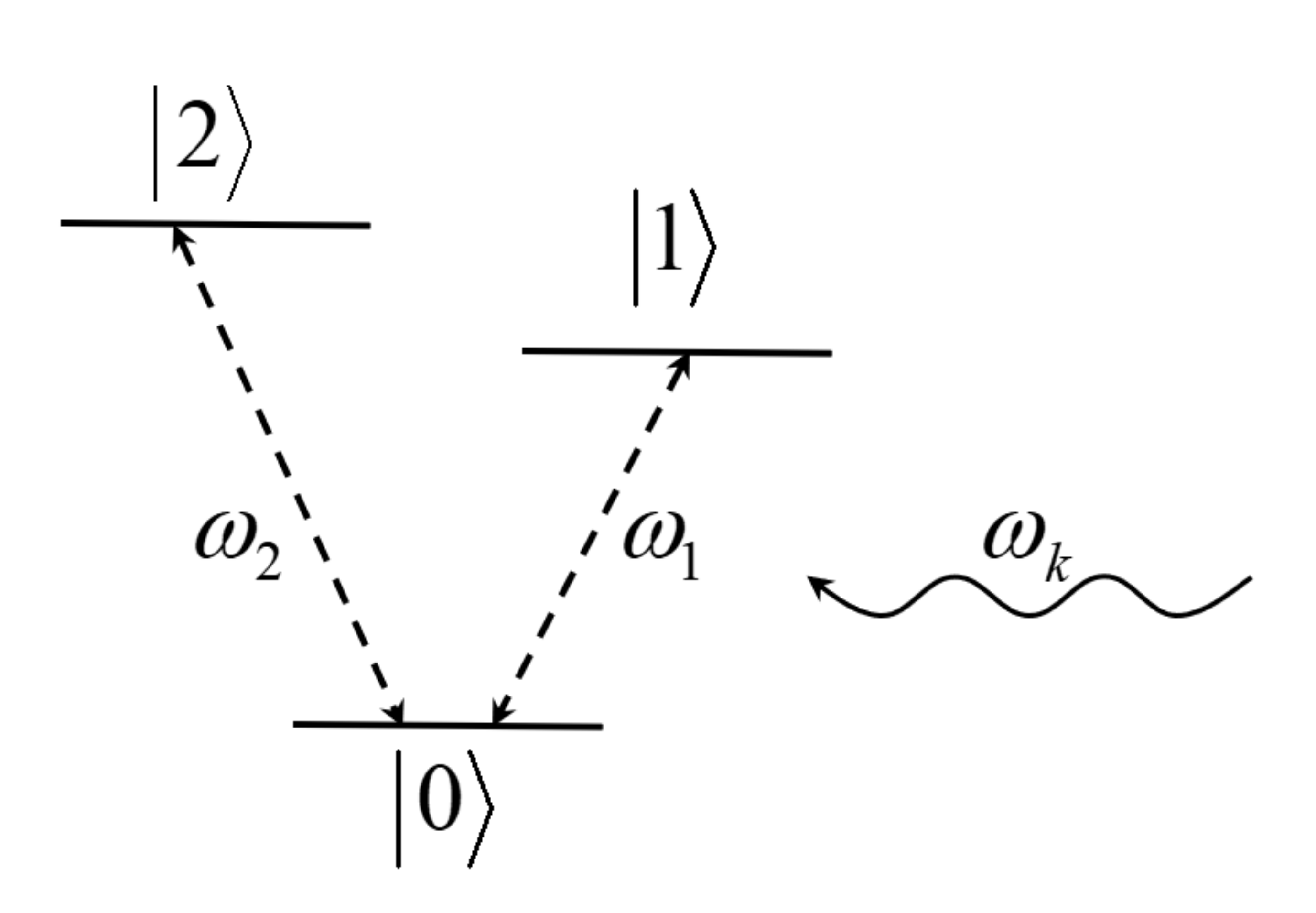}
\caption{the V-type atom}
\label{fig:1}      
\end{center}
\end{figure}
In this atom, the transition $|0\rangle\rightleftharpoons1\rangle$ and $|0\rangle\rightleftharpoons|2\rangle$ are allowed, whereas the transition $|1\rangle\rightleftharpoons|2\rangle$ is forbidden in the electric-dipole approximation, as shown in figure.\ref{fig:1}; the atomic transition frequency is $\omega_{1}$ between atomic states $|0\rangle$ and $|1\rangle$ and $\omega_{2}$ between atomic states $|0\rangle$ and $|2\rangle$. The Hamiltonian describing the dynamics of the system in the rotating wave approximation can be written as ($\hbar=1$):
\begin{eqnarray}
   \hat{H}=\hat{H}_{0}+\hat{H}_{I},
\end{eqnarray}
where
\begin{eqnarray}
       \hat{H}_{0}=\omega_{1}|1\rangle\langle1|+\omega_{2}|2\rangle\langle2|+\sum_{k}\omega_{k}\hat{a}_{k}^{\dagger}\hat{a}_{k},
\end{eqnarray}
and
\begin{eqnarray}
   \hat{H}_{I}=\sum_{k}(g_{k_{1}}\hat{a}_{k}^{\dagger}|0\rangle\langle1|+g_{k_{2}}\hat{a}_{k}^{\dagger}|0\rangle\langle2|+H.c.),
\end{eqnarray}
where $\hat{a}_{k}$ and $\hat{a}_{k}^{\dagger}$ are respectively the annihilation and creation operators of the k-th reservoir, with the k-th field mode frequency $\omega_{k}$ in the reservoir. $g_{k_{1}}$represents the coupling constants between the $|0\rangle\rightleftharpoons|1\rangle$ transition and the k-th reservoir.
Likewise, the coupling constants between the $|0\rangle\rightleftharpoons|2\rangle$ transition and the k-th reservoir is $g_{k_{2}}$.

Considering the atomic transition frequency $\omega_{1}=\omega_{2}$, the evolution of the atom subsystem can be obtained by Kraus operators\cite{N.2017}
\begin{eqnarray}
   \rho_s(t)=\sum_{i=1}^{3}K_i\rho_s(0)K_i,
\end{eqnarray}
where
\begin{eqnarray}
   K_i=U^{\dag}\mathcal {K}_iU (i\in\{1,2,3\}),
\end{eqnarray}
and
\begin{eqnarray}
   \mathcal {K}_1&=&\left(
                   \begin{array}{cccc}
                    G_+(t) & 0&$ $ 0\\
                    0 &  G_-(t) &$ $ 0\\
                    0 &  0  &   $ $  1\\
                    \end{array}
              \right),
\nonumber\\
\nonumber\\
   \mathcal {K}_2&=&\left(
                   \begin{array}{cccc}
                    0 &$ $  0 & $ $0\\
                    0 &$ $   0 &$ $0\\
                    \sqrt{1-|G_+(t)|^2} &$ $  0 &$ $ 0\\
                    \end{array}
              \right),
\nonumber\\
\nonumber\\
   \mathcal {K}_3&=&\left(
                   \begin{array}{cccc}
                    0 &$ $  0 & $ $0\\
                    0 &$ $  0 & $ $0\\
                    0 &\sqrt{1-|G_-(t)|^2} &$ $ 0\\
                    \end{array}
              \right).
\end{eqnarray}
In this equation, $G_{\pm}(t)$ is given by
\begin{small}
\begin{eqnarray}
G_{\pm}(t)=e^{\frac{-\lambda t}{2}}\left[\cosh \left( \frac{d_{\pm}t}{2}\right)+\frac{\lambda}{d_{\pm}}\sinh\left(\frac{d_{\pm}t}{2}\right)\right]
\end{eqnarray}
\end{small}with $d_{\pm}=\sqrt{\lambda^2-2\lambda\gamma_{\pm}}$, $\gamma_{\pm}=\frac{\gamma_1+\gamma_2\pm q}{2}$ and $q=\sqrt{(\gamma_1-\gamma_2)^2+4\gamma_{12}^2}$.
Where $\lambda$ is the spectral width of the distribution.
When the $\lambda\gg\gamma_{\pm}$, the behavior of the system is Markovian; when $\lambda<2\gamma_{\pm}$, it means the strong coupling
regime and the behavior of the system is non-Markovian.
$\gamma_i$ is the spontaneous decay constant of the excited sublevel $i$ to the ground level $|0\rangle$ and $\gamma_{ij}=\sqrt{\gamma_i\gamma_j}\theta$ $(i\neq j$ and $|\theta|\leq1)$ represents the effect of quantum interference resulting from the cross coupling between the transitions $|2\rangle\rightarrow|0\rangle$ and $|1\rangle\rightarrow|0\rangle$. $\theta$ depends on the relative angle between two dipole moment element related to the mentioned transitions. $\theta=0$ means that the dipole moments of two transitions are perpendicular to each other corresponding to the case that there is no the spontaneously generated interference (SGI) between two decay channels. On the other hand, $\theta=\pm1$ indicate that the two dipole moments are parallel or antiparallel corresponding to the interference effect between two decay channels is maximal.
In the equation (8), the unitary transformation U is given by:
\begin{eqnarray}
U&=&\left(
                   \begin{array}{cccc}
                    A &-B & $ $0\\
                    B & A& $ $0\\
                    0 &0 &$ $ 1\\
                    \end{array}
              \right),
\end{eqnarray}
where $A=\sqrt{\frac{q+\gamma_1-\gamma_2}{2q}};B=\sqrt{\frac{q-\gamma_1+\gamma_2}{2q}}$.
And then, using the unitary transformation U, we get the following expressions for $K_i$:
\begin{small}
\begin{eqnarray}
  \fl  K_1&=&\left(
                   \begin{array}{cccc}
                    G_+A^2+G_-B^2& (-G_++G_-)AB&  0\\
                    (-G_++G_-)AB& G_+B^2+G_-A^2&  0\\
                    0 &  0  &  1\\
                    \end{array}
              \right),
\nonumber\\
\nonumber\\
   K_2&=&\sqrt{(1-|G_+|^2)}\left(
                   \begin{array}{cccc}
                    0 &$ $  0 & $ $0\\
                    0 &$ $   0 &$ $0\\
                    \sqrt{A} &$ $-\sqrt{B} &$ $ 0\\
                    \end{array}
              \right),
\nonumber\\
\nonumber\\
   K_3&=&\sqrt{(1-|G_-|^2)}\left(
                   \begin{array}{cccc}
                    0 &$ $  0 & $ $0\\
                    0 &$ $  0 & $ $0\\
                    \sqrt{B} &$\ \  $\sqrt{A}&$ $ 0\\
                    \end{array}
              \right),
\end{eqnarray}
\end{small}
hence, it is concluded that $\sum_{i=1}^3K_i^\dag K_i=I_3$ and $\rho_s(t)=\sum_{i=1}^3K_i\rho_s(0)K_i^\dag$.

In this paper, we consider the relation rates of two decay channels are equal, $\gamma_1=\gamma_2=\gamma$.
\section{Reduction of the entropy uncertainty for qutrit with a memory system }

Now let us consider that Bob initially prepares a pair of qutrits A and B being entangled, as follows
\begin{eqnarray}
\rho_k(0)=\frac{1-k}{9}I+k|\psi_+\rangle\langle\psi_+|,
\end{eqnarray}
in this equation, $|\psi_+\rangle=\frac{1}{\sqrt{3}}(|00\rangle+|11\rangle+|22\rangle)$ is the maximally entangled pure state.
Bob sends qutrit A  to Alice who measures either R or Q on qutrit A and informs Bob of her choice.
Meanwhile Bob serves qutrit B as quantum memory.
In this case, two qutrits each part locally interacts with a independent reservoir.
The evolution of a two-qutrit state based on the techniques of Kraus can be characterized as\cite{Bellomo.2007}
\begin{eqnarray}
\rho_{AB}(t)=\sum_{i=1,j=1}^{3}(K_i\otimes K_j)\rho_k(0)(K_i\otimes K_j)^{\dagger},
\end{eqnarray}
where $K_i$ is Kraus operator.
In this section, we choose the spin-1 observable $R(Q)=S_x(S_z)$.
The conditional von-Neumann entropy after qutrit A was measured by Alice($S_x$ or $S_z$) is given as
\begin{eqnarray}
S(S_x|B)=S(\rho_{S_xB})-S(\rho_B),\nonumber\\
S(S_z|B)=S(\rho_{S_zB})-S(\rho_B),
\end{eqnarray}
the post measurement state
$\rho_{XB}=\sum_i(|\psi_i\rangle\langle\psi_i|\otimes I)\rho_{AB}(|\psi_i\rangle\langle\psi_i|\otimes I)$, where ${|\psi_i\rangle}$ is the eigenstate of the observable $X$ and $I$ is an identity operator in the Hilbert space of particle B.
$\rho_B=tr_A(\rho_{AB(t)})$ is the density matrix of subsystem B. So the left-hand  side of equation(3) can be given by
\begin{eqnarray}
U_L=S(S_x|B)+S(S_z|B),
\end{eqnarray}
and the right-hand side of equation(3) can be represented as
\begin{eqnarray}
U_b=\log_{2}\frac{1}{c}+S(\rho_{AB})-S(\rho_B).
\end{eqnarray}
$S(\rho)=\sum_i\lambda_i\log_2\lambda_i$ is the von Neumann entropy where $\lambda_i$ is the eigenvalue of $\rho$.

In the traditional the entropy uncertainty relation equation(2), if we choose the spin-1 observable $R(Q)=S_x(S_z)$, the limit of the traditional lower bound of the entropy uncertainty can be calculated
\begin{eqnarray}
U_b=\log_{2}\frac{1}{c}=\log_{2}\frac{1}{2}=1
\end{eqnarray}
The degree of entanglement for qutrits system can be easily computed by the negativity $\mathcal{N}$ which is given by
\begin{eqnarray}
\mathcal{N}=\frac{\parallel\rho^{T_A}\parallel-1}{2}.
\end{eqnarray}
The matrix $\rho^{T_A}$ is the partial transpose with respect to the subsystem A, that is $\rho_{ik,jl}^{T_A}=\rho_{jk,il}$, and the trace norm $\parallel\cdot\parallel$ is defined as $\parallel\rho\parallel=tr(\rho\rho^{\dagger})^{1/2}$.
\begin{figure}[htpb]
\begin{center}
  \includegraphics[width=0.4\textwidth]{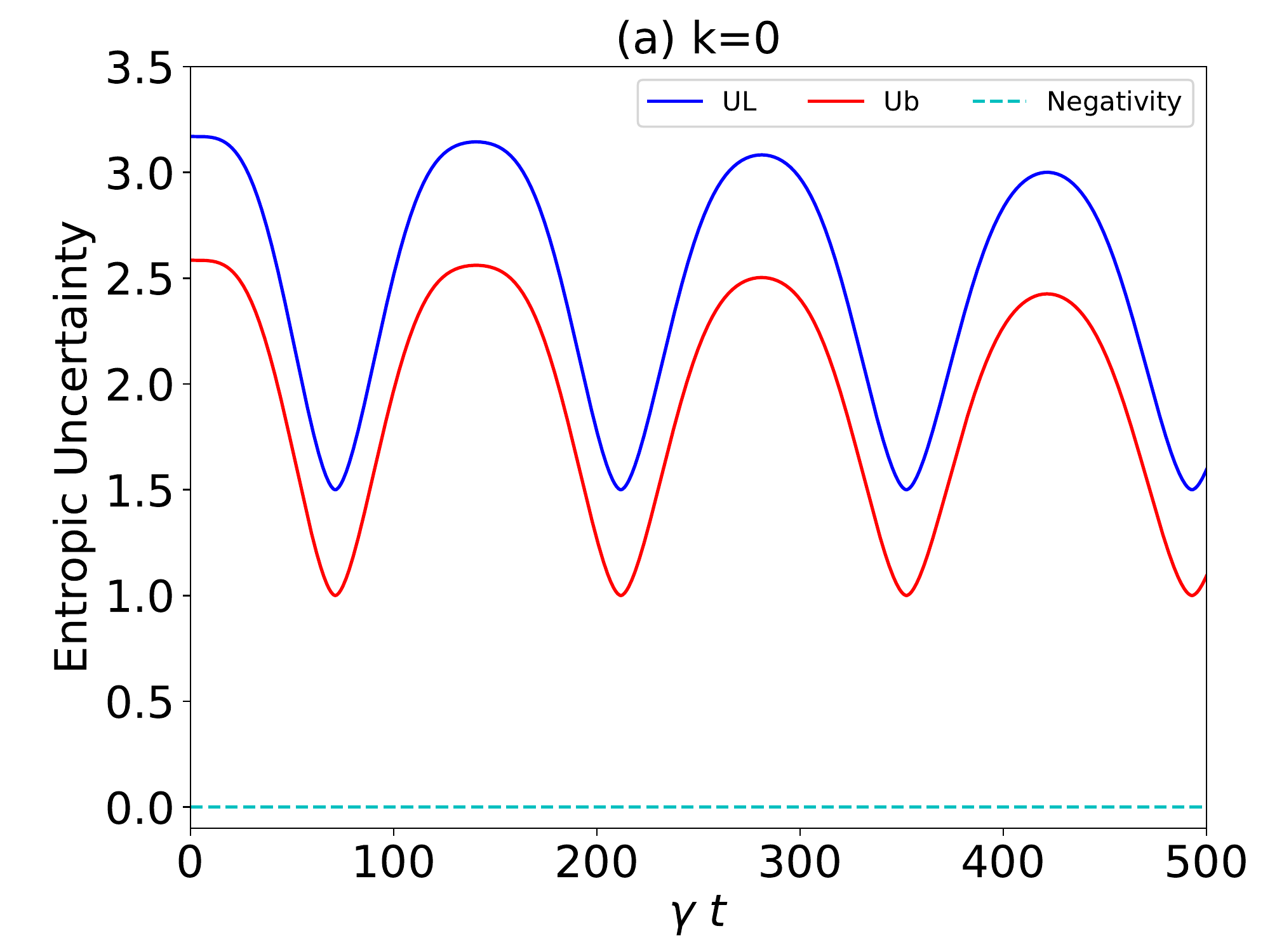}
  \includegraphics[width=0.4\textwidth]{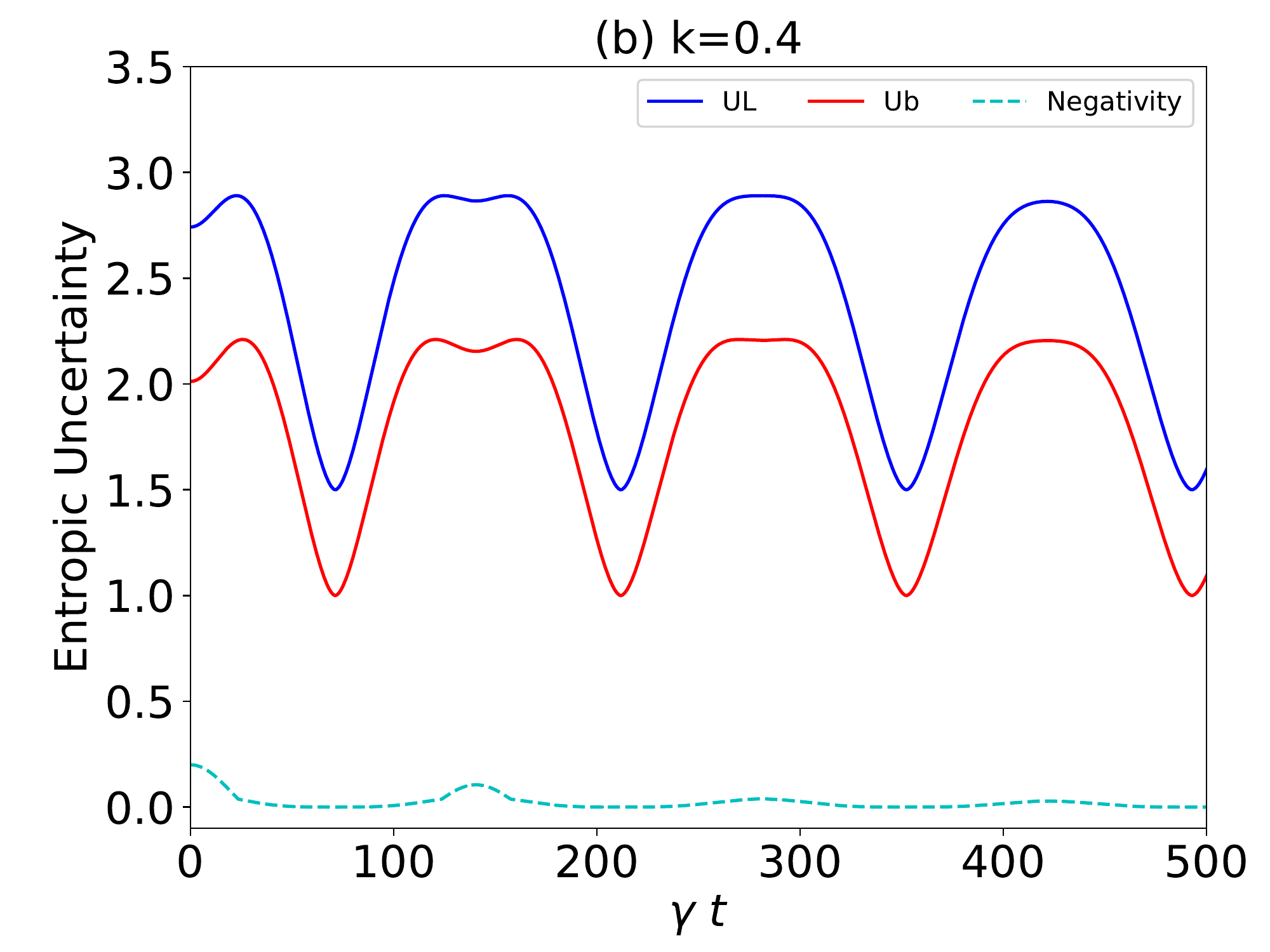}
  \includegraphics[width=0.4\textwidth]{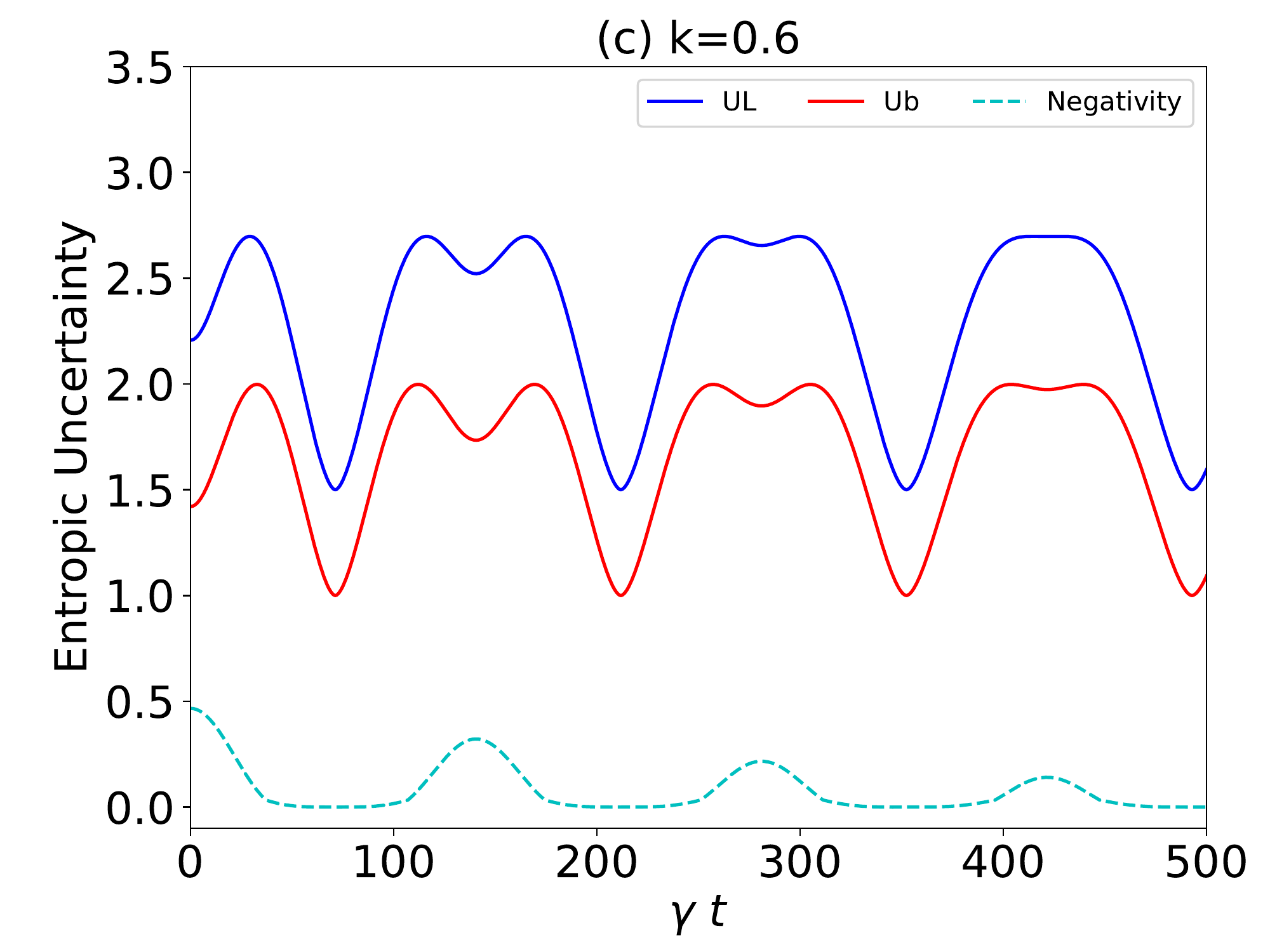}
  \includegraphics[width=0.4\textwidth]{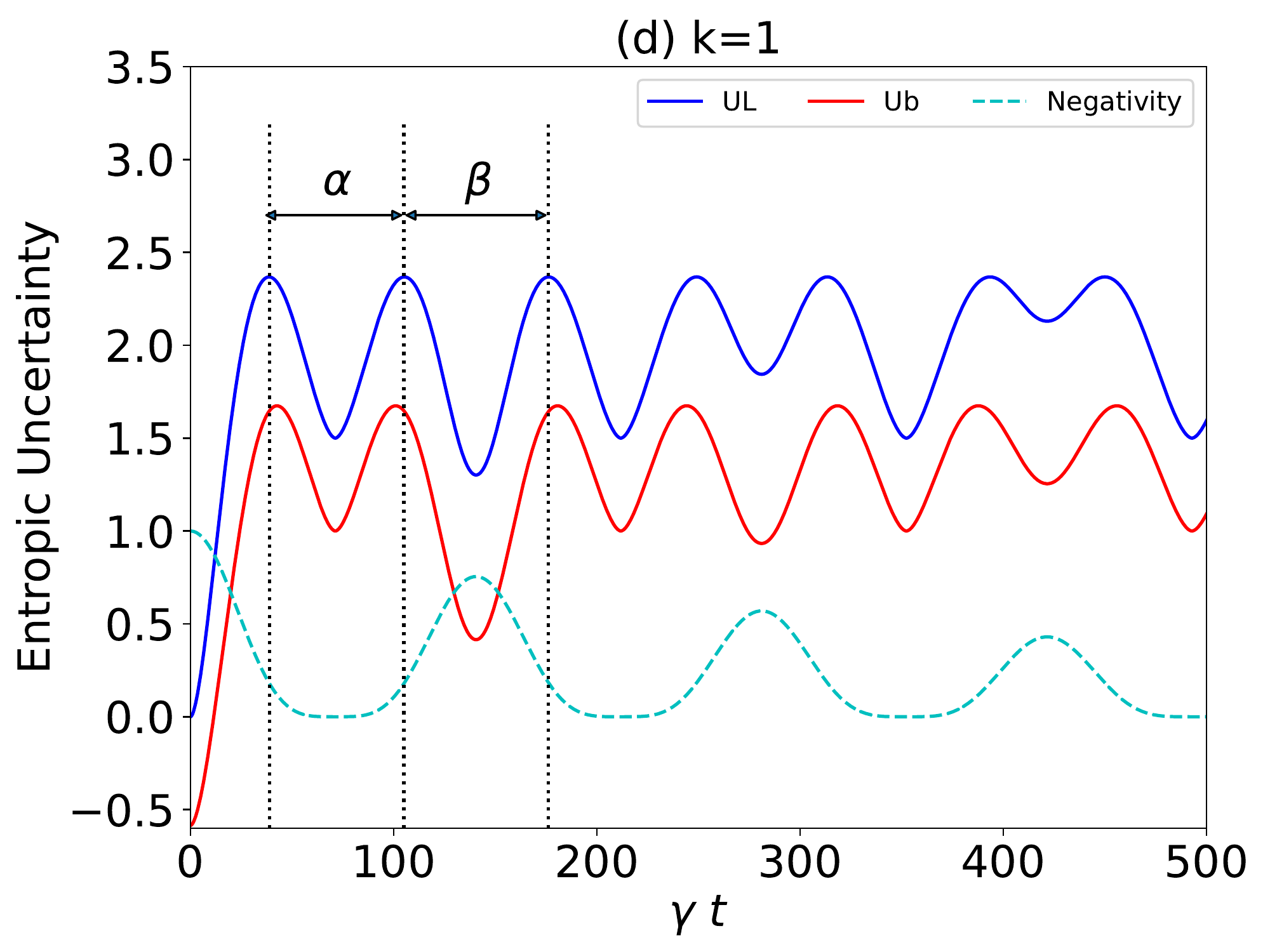}
\caption{The evolution of the entropy uncertainty (the blue solid  line), its lower bound (the red solid line) and the negativity (the cyan dotted line) in no SGI($\theta=0$) and $\lambda=1000\gamma$ case considering the following parameter of initial state: (a) $k=0$; (b) $k=0.4$;(c) $k=0.6$;(d) $k=1$.}
\label{fig:2}      
\end{center}
\end{figure}

In order to study the dynamics of the entropy uncertainty for qutrits system in a non-markov environment, we draw the change curve of the entropy uncertainty, its lower bound and the negativity $\mathcal{N}$  with time in figure.\ref{fig:2}, both of them considering the strong non-markov case $\lambda=1000\gamma$ .
In figure.\ref{fig:2}(a), we choose initial state as $\rho_{k=0}(0)$ which means that two qutrits are free of entanglement and the system has no memory effects.
From the figure.\ref{fig:2}(a), on the one hand, one can see the negativity is always 0 because two qutrits are free of entanglement.
On the other hand, it can be seen that that the entropy uncertainty $U_L, U_b$ exhibit oscillations with the evolutionary period of $\gamma t\approx149$.
If we select the parameter of initial state as $k=0.4$, $k=0.6$ and $k=1$, shown as figure.\ref{fig:2}(b), (c) and (d), there are memory effect.
 It is find that the entropy uncertainty and its lower limit are reduced  with memory effect, by comparing figure.\ref{fig:2}(a)with(b),(c),(d).
 The stronger strength of the initial state entanglement is, the more the entropy uncertainty reduces: from figure.\ref{fig:2}(a) to (d), the maximum of the entropy uncertainty $3.169\rightarrow 2.890\rightarrow 2.697\rightarrow 2.368$.
 In addition, the initial entropy uncertainty is 0 when the initial state is max entanglement which breaks the limit of the traditional lower bound of the entropy uncertainty seeing figure.\ref{fig:2}(d) .

Figure.\ref{fig:2}(d) also reveals that, the evolutionary period of entropy uncertainty can be divided into two regions: $\alpha$ and $\beta$ region.
Our result is different from the conventional wisdom that the entropy uncertainty is negatively correlated with entanglement.
In the $\alpha$ region the entropy uncertainty decreases and then slowly increases meanwhile the negativity decays asymptotically to 0 and after a freeze  time, the negativity began to increase.
In the $\beta$ region, as the time goes the entropy uncertainty decreases with the negativity increasing and then the entropy uncertainty increases with the negativity decreasing.
In other word, the entropy uncertainty is negatively correlated with entanglement in the $\beta$ region, not in the $\alpha$ region.
Beside in the $\beta$ region, the minimum value of the entropy uncertainty can be reduce more.
These results show that the memory systems formed by two qutrits entangled reduce effectively the entropy uncertainty and break the traditional lower bound of the entropy uncertainty when degree of the entanglement is large.
 In the non-Markov environment, entanglement only affect the the entropy uncertainty locally and cannot completely control the evolution of the entropy uncertainty. In the $\beta$ region, entanglement is beneficial to reduce entropy uncertainty.
 \section{Reduction of the entropy uncertainty for qutrit with SGI }
\begin{figure}[htpb]
\begin{center}
  \includegraphics[width=0.4\textwidth]{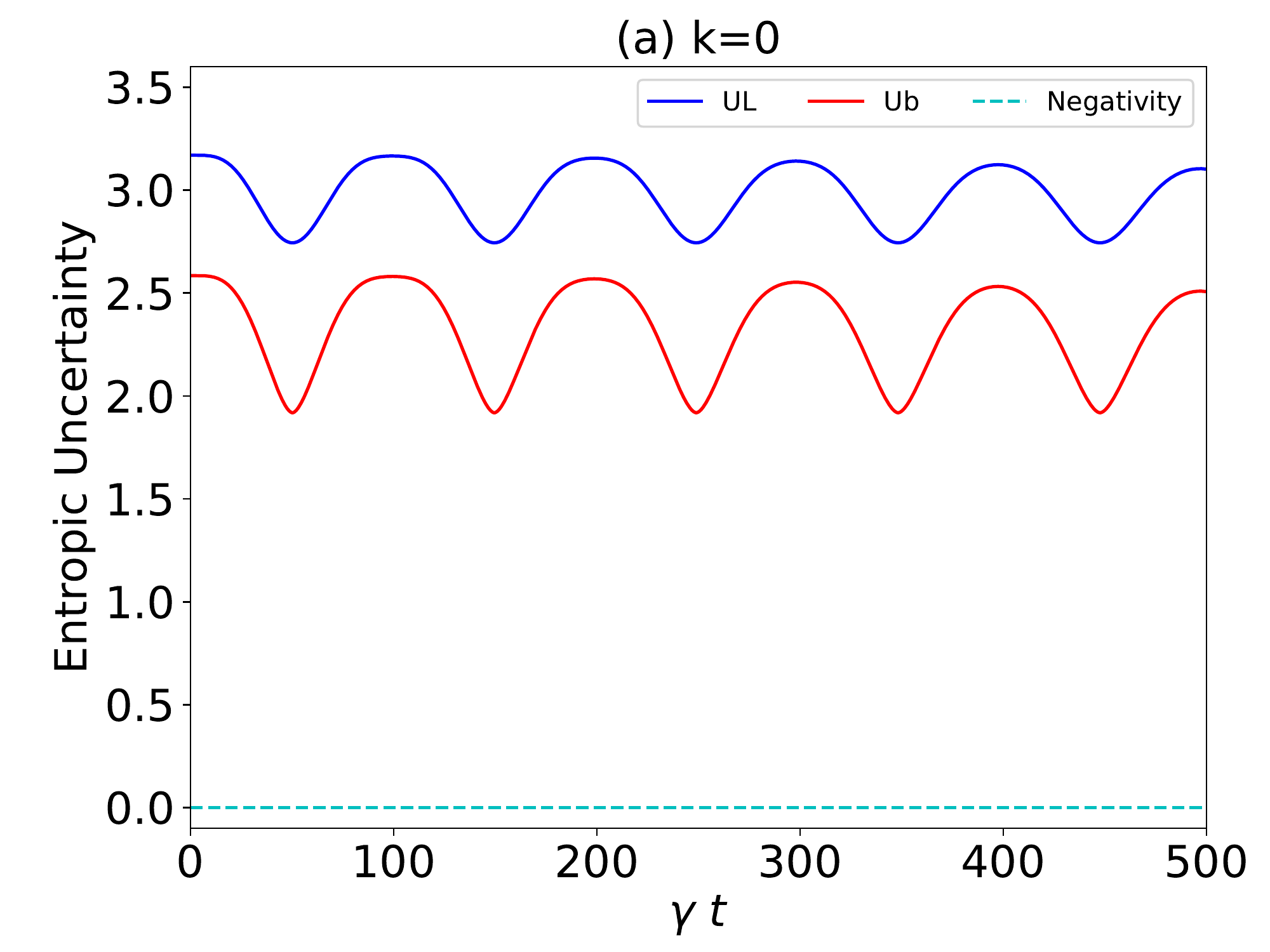}
  \includegraphics[width=0.4\textwidth]{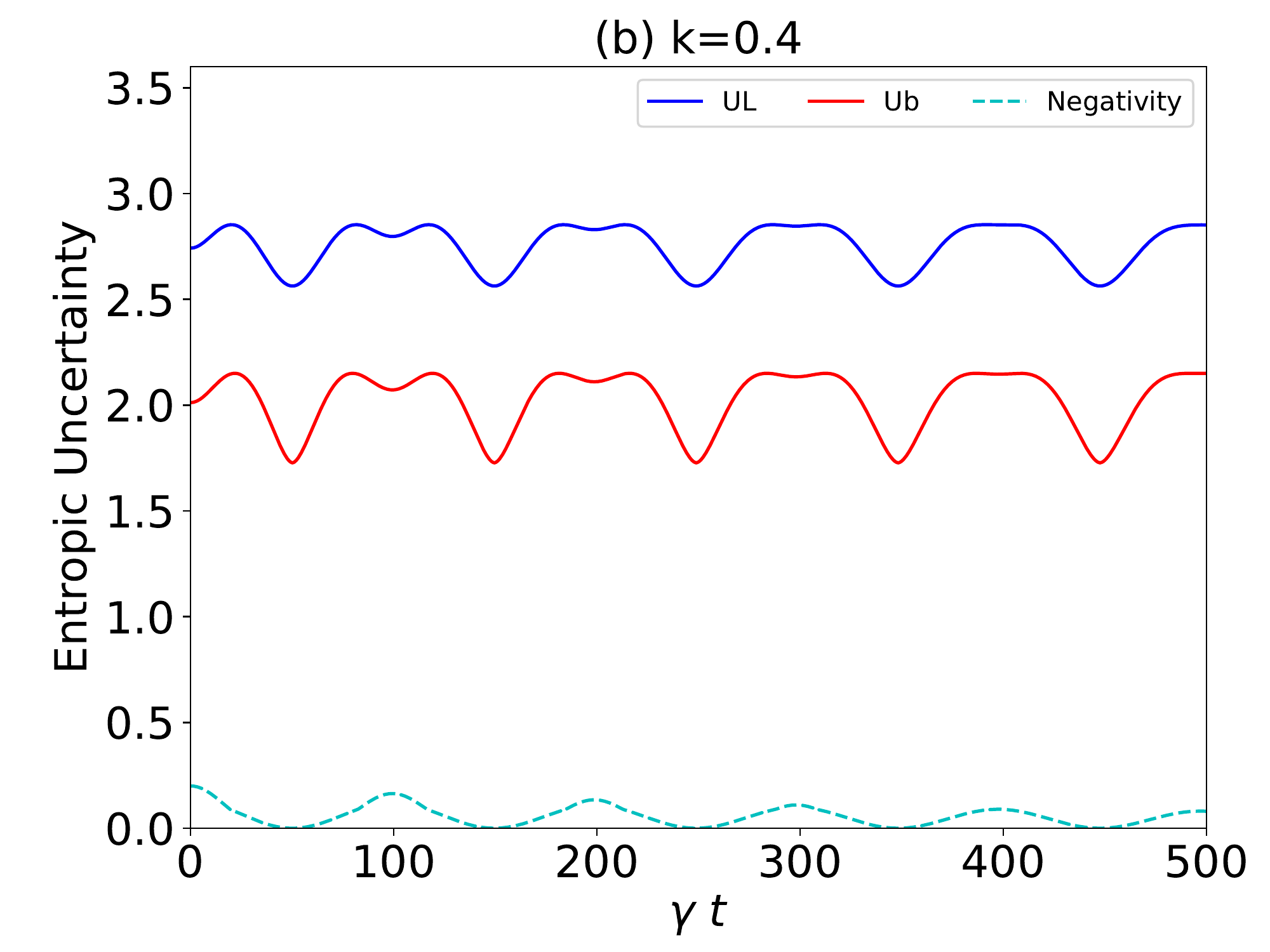}
  \includegraphics[width=0.4\textwidth]{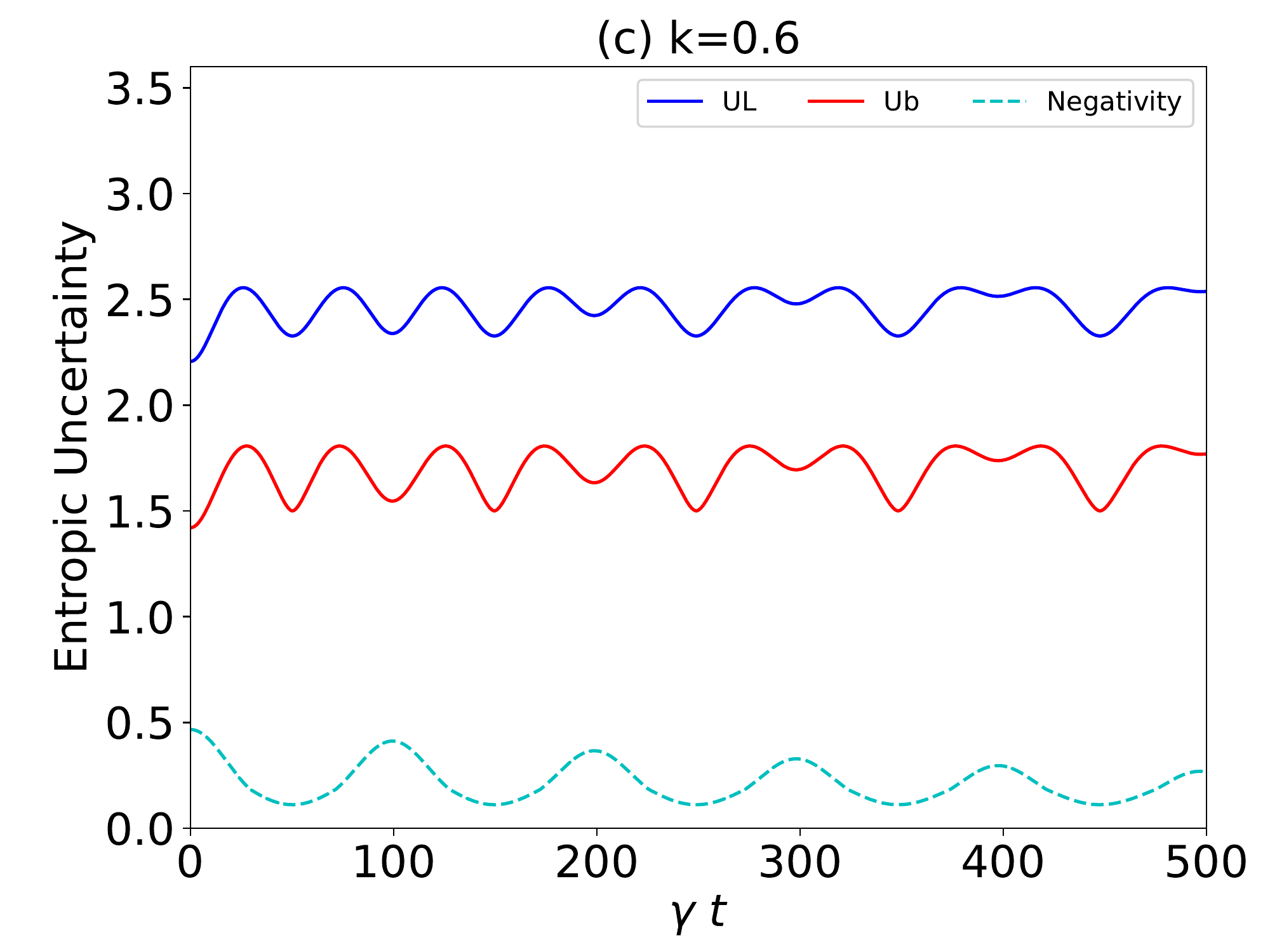}
  \includegraphics[width=0.4\textwidth]{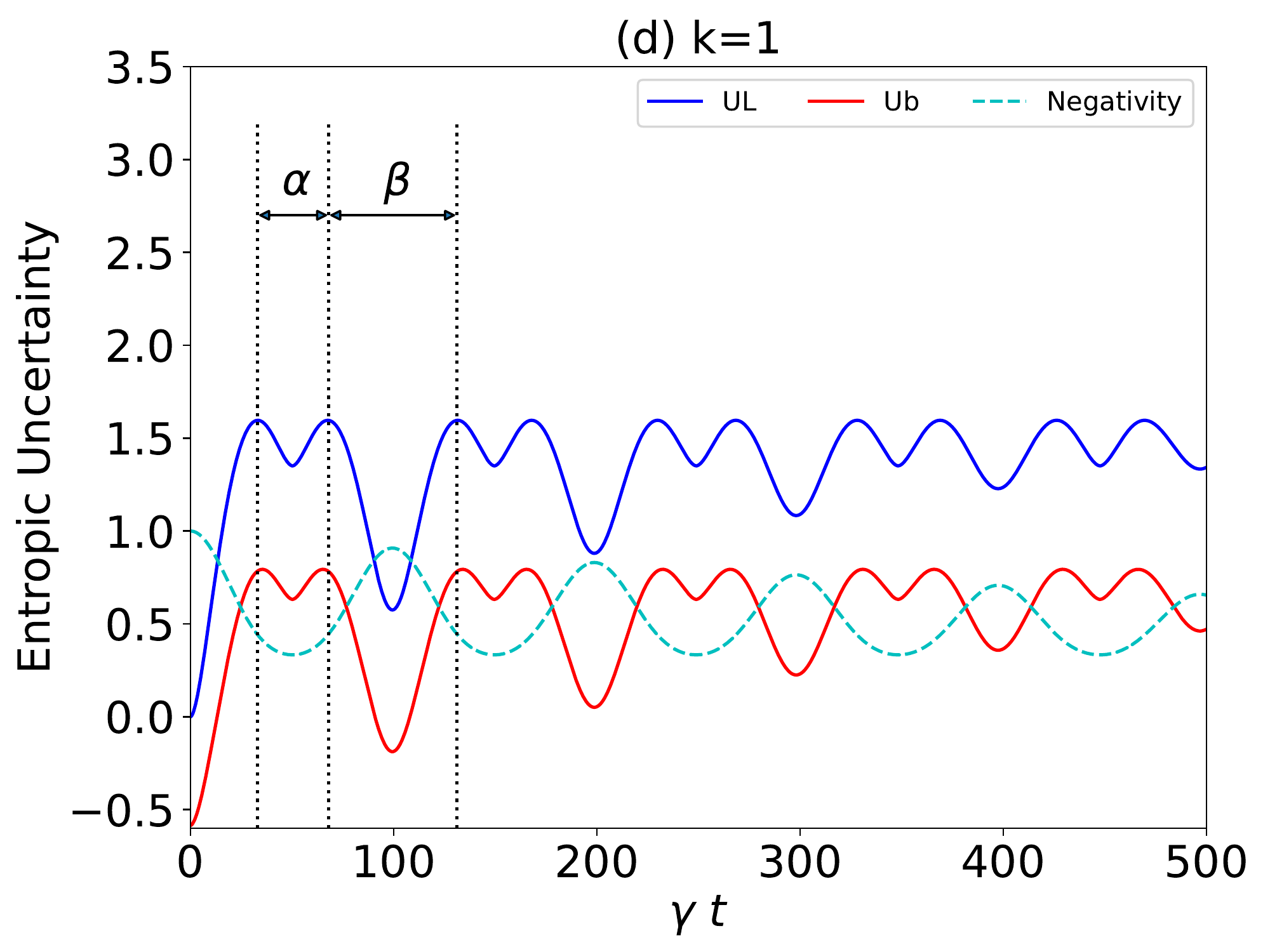}
\caption{The evolution of the the entropy uncertainty (the blue solid line), its lower bound (the red solid line) and the negativity (the cyan dotted line) in strongest SGI($\theta=1$) and $\lambda=1000\gamma$ case considering the following parameter of initial state: (a) $k=0$; (b) $k=0.4$;(c) $k=0.6$;(d) $k=1$.}
\label{fig:3}      
\end{center}
\end{figure}
So far, we have only discussed the case of no SGI. Next we will study the case of strongest SGI which means the two dipole moment elements of qutrit are parallel (or antiparallel).
figure.\ref{fig:3} presents the numerical results of the entropy uncertainty, its lower bounds and negativity in $\lambda=1000\gamma$, when the two dipole moment elements of qutrit are parallel.
In figure.\ref{fig:3}(a), we have assumed the initial state parameters $k=0$. The dynamics of the entropy uncertainty is similar to that found in the no-SGI case in figure.\ref{fig:2}(a), except for the evolutionary period.
One can see that in the no memory effects case, the SGI effect cannot reduce the entropy uncertainty.
But if we consider the memory effects, the entropy uncertainty can be further reduced by the SGI effect. The maximum of the entropy uncertainty from  2.890 to 2.853 in the $k=0.4$; from 2.697 to 2.555 in the $k=0.6$ and from 2.368 to 1.596 in the $k=1$ are revealed in figure. \ref{fig:3}(b), (c) and (d).
It is clearly seen that the stronger strength of the initial state entanglement is, the SGI effect can reduce the entropy uncertainty more.
In figure.\ref{fig:3}(d), it should be noted that in the first $\beta$ region, the minimum value of the entropy uncertainty is $0.575<1$.
It  is lower than the limit of the traditional lower bound of the entropy uncertainty which is not found in figure.\ref{fig:2}(d).
These results indicates that the strongest SGI can effectively reduce the entropy uncertainty in a two qutrit entangled system.
\begin{figure}[htpb]
\begin{center}
  \includegraphics[width=0.4\textwidth]{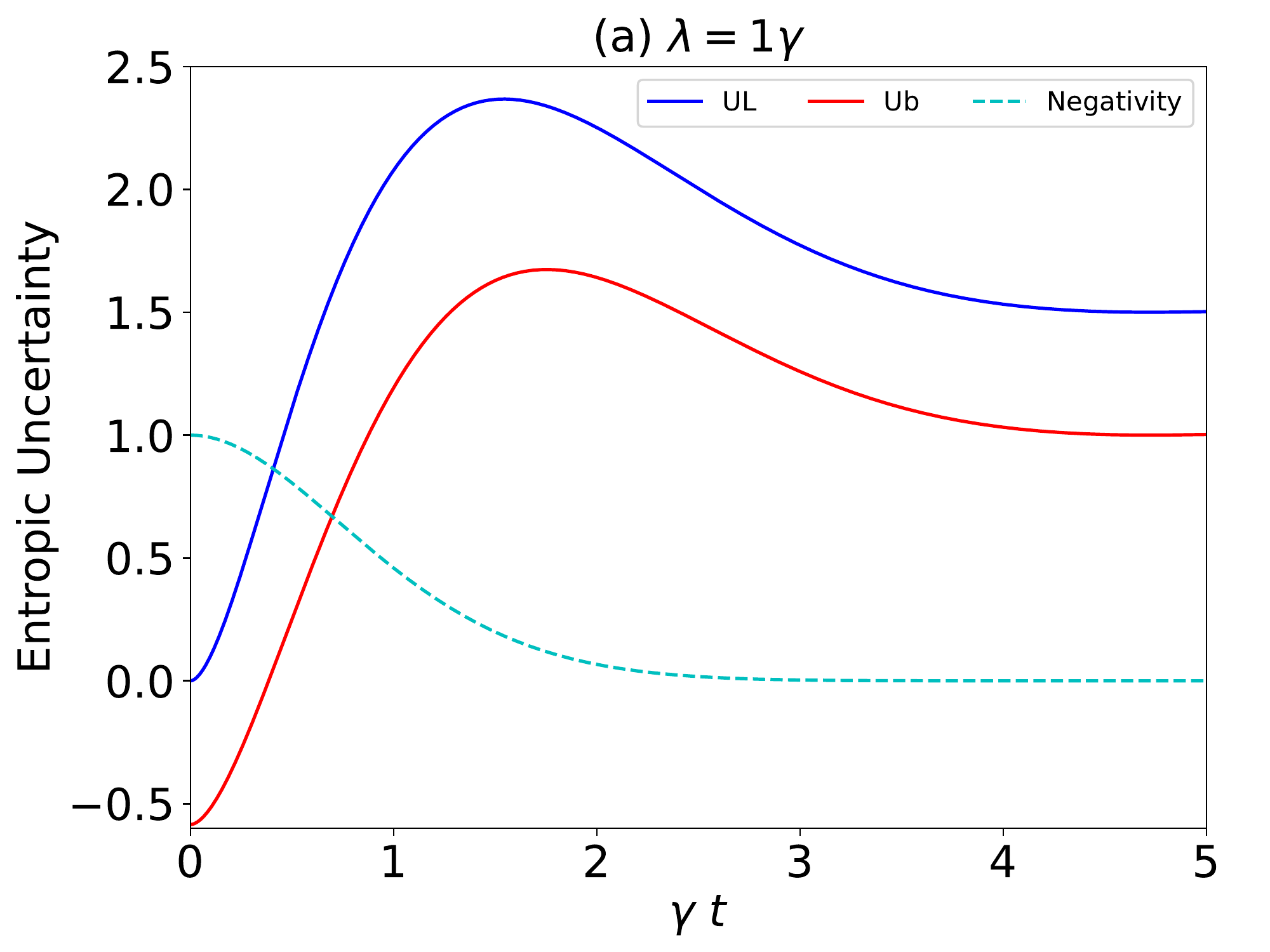}
  \includegraphics[width=0.4\textwidth]{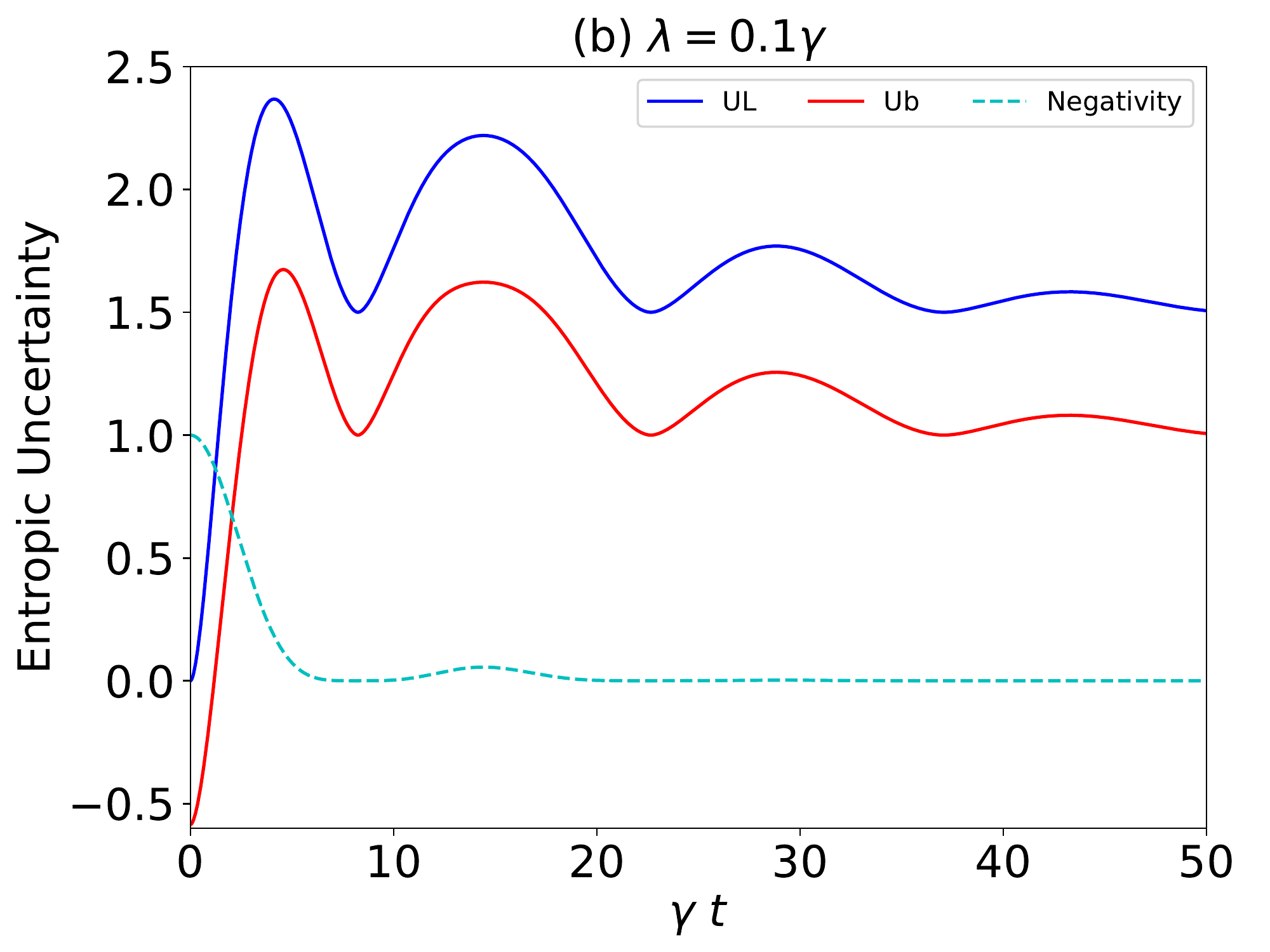}
  \includegraphics[width=0.4\textwidth]{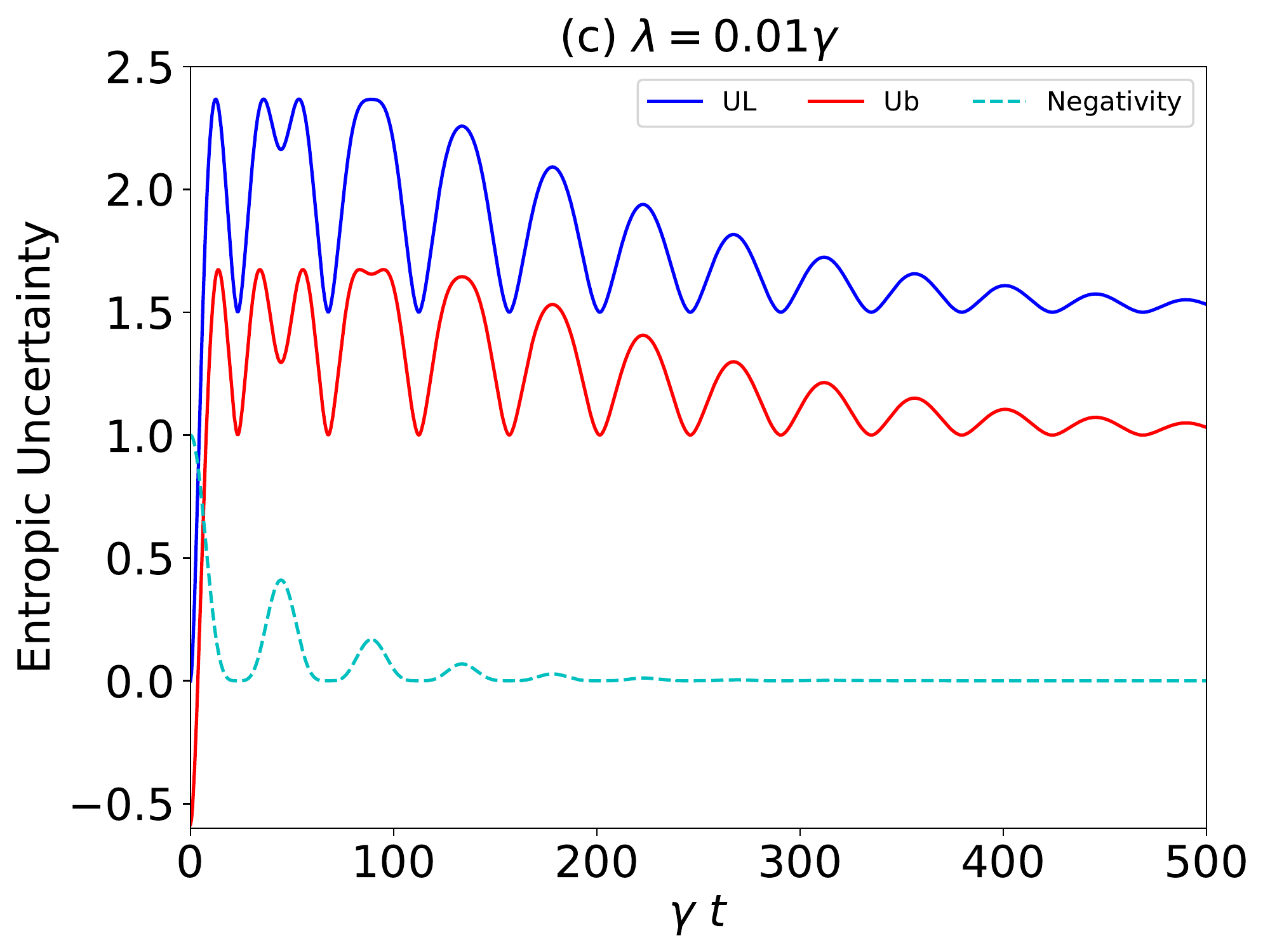}
  \includegraphics[width=0.4\textwidth]{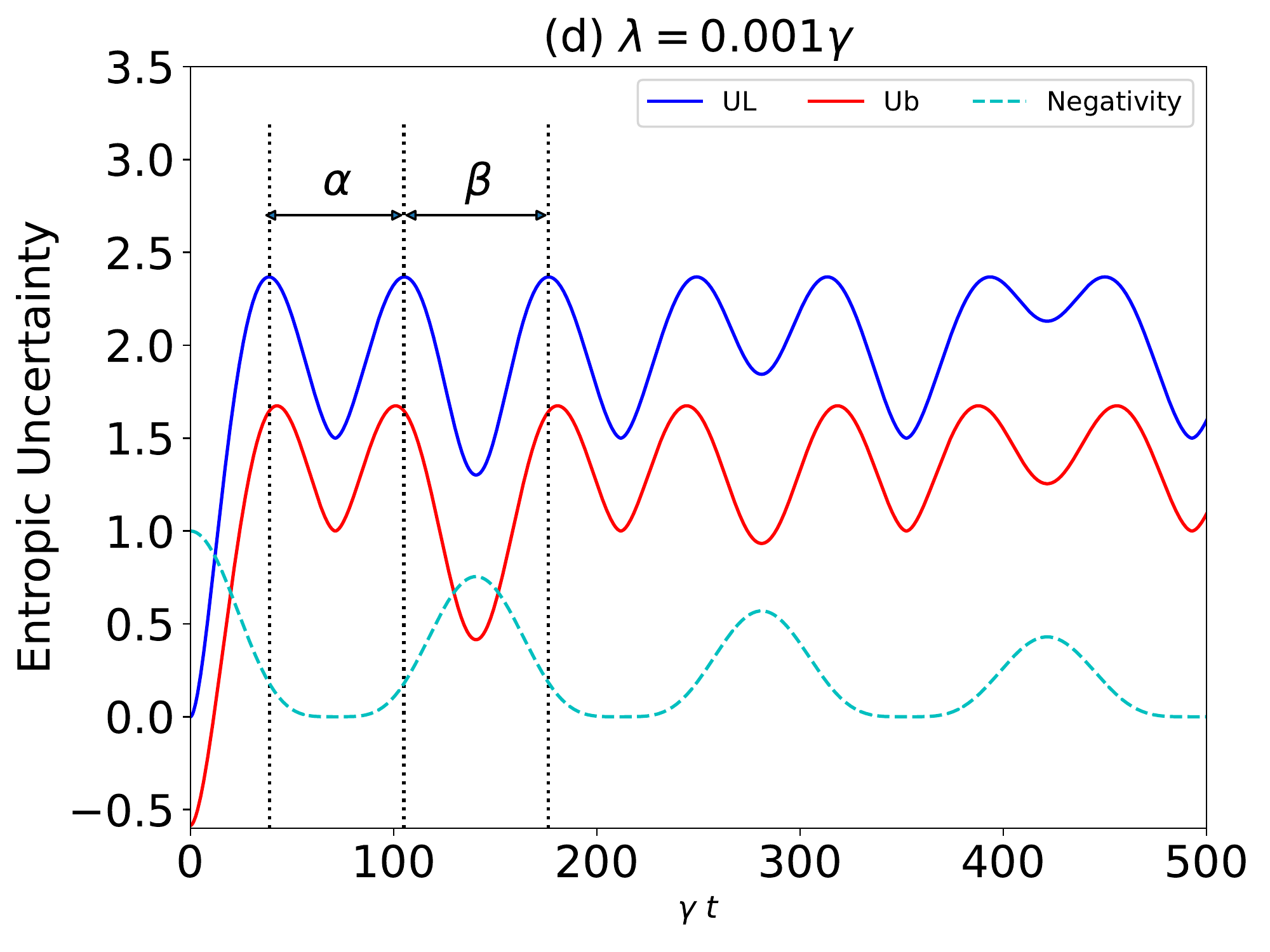}
 \end{center}
\caption{TThe evolution of the the entropy uncertainty (the blue solid line), its lower bound (the red solid line) and the negativity (the cyan dotted line)  in no SGI($\theta=0$) and considering the initial state maximum entangled for different non-markov strengths: (a) $\lambda=\gamma$; (b)$\lambda=0.1\gamma$ ;(c)$\lambda=0.01\gamma$; (d)$\lambda=0.001\gamma$.}
\label{fig:4}      
\end{figure}

Finally, influence of non-markov strengths on the entropy uncertainty is discussed. Figure.\ref{fig:4} shows our findings considering the initial state maximum entangled in the no SGI case for different non-markov strengths. From figure.\ref{fig:4} one can see that with the non-markov strengths increasing, the max value of the entropy uncertainty is not reduced, but the oscillation frequency of the entropy uncertainty increases. It is note that with the non-markov strengths increasing,the evolutionary period of entropy uncertainty can be divided into two regions: $\alpha$ and $\beta$ region, discussed in section 3. In other word, the non-markov effect can reduces the local minimum of the entropy uncertainty but does not affect the maximum of the entropy uncertainty.

\section{Conclusion}
In conclusion, we have investigated the dynamics of the entropy uncertainty for a qutrit with and without a memory system under non-Markov noisy environment in two cases: no-SGI and the strongest SGI.
We have found that the entropy uncertainty is reduced effectively with a memory system
and the stronger strength of the initial state entanglement is, the more the entropy uncertainty reduces.
The traditional lower bound of uncertainty is broken when degree of the entanglement is large.
Furthermore, in the strongest SGI case the entropy uncertainty can be further reduced for two-qutrit entangled system.
 The stronger strength of the initial state entanglement is, the SGI effect can reduce the entropy uncertainty more.
To sum up, our investigation might offer an insight into the dynamics of the entropy uncertainty in a realistic system, and be nontrivial to quantum precision measurement in prospective quantum information processing.

\section*{Acknowledgments}
This work is supported by the National Natural Science Foundation of China (Grant No.11374096).
\section*{References}
\bibliographystyle{unsrt}
\bibliography{paper1}
\end{document}